\input harvmac.tex
%
%
%
\message{S-Tables Macro v1.0, ACS, TAMU (RANHELP@VENUS.TAMU.EDU)}
%
%
\newhelp\stablestylehelp{You must choose a style between 0 and 3.}%
\newhelp\stablelinehelp{You should not use special hrules when stretching
a table.}%
\newhelp\stablesmultiplehelp{You have tried to place an S-Table inside another
S-Table.  I would recommend not going on.}%
%
%
\newdimen\stablesthinline
\stablesthinline=0.4pt
\newdimen\stablesthickline
\stablesthickline=1pt
%
%
\newif\ifstablesborderthin
\stablesborderthinfalse
\newif\ifstablesinternalthin
\stablesinternalthintrue
\newif\ifstablesomit
\newif\ifstablemode
\newif\ifstablesright
\stablesrightfalse
%
%
\newdimen\stablesbaselineskip
\newdimen\stableslineskip
\newdimen\stableslineskiplimit
%
%
\newcount\stablesmode
\newcount\stableslines
\newcount\stablestemp
\stablestemp=3
\newcount\stablescount
\stablescount=0
\newcount\stableslinet
\stableslinet=0
%
%
%
\newcount\stablestyle
\stablestyle=0
%
%
\def\stablesleft{\quad\hfil}%
\def\stablesright{\hfil\quad}%
%
%
\catcode`\|=\active%
%
%
\newcount\stablestrutsize
\newbox\stablestrutbox
\setbox\stablestrutbox=\hbox{\vrule height10pt depth5pt width0pt}
\def\stablestrut{\relax\ifmmode%
                         \copy\stablestrutbox%
                       \else%
                         \unhcopy\stablestrutbox%
                       \fi}%
%
%
\newdimen\stablesborderwidth
\newdimen\stablesinternalwidth
\newdimen\stablesdummy
\newcount\stablesdummyc
\newif\ifstablesin
\stablesinfalse
%
%
\def\begintable{\stablestart%
  \stablemodetrue%
  \stablesadj%
  \halign%
  \stablesdef}%
\def\stablesadj{%
  \ifcase\stablestyle%
    \hbox to \hsize\bgroup\hss\vbox\bgroup%
  \or%
    \hbox to \hsize\bgroup\vbox\bgroup%
  \or%
    \hbox to \hsize\bgroup\hss\vbox\bgroup%
  \or%
    \hbox\bgroup\vbox\bgroup%
  \else%
    \errhelp=\stablestylehelp%
    \errmessage{Invalid style selected, using default}%
    \hbox to \hsize\bgroup\hss\vbox\bgroup%
  \fi}%
\def\stablesend{\egroup%
  \ifcase\stablestyle%
    \hss\egroup%
  \or%
    \hss\egroup%
  \or%
    \egroup%
  \or%
    \egroup%
  \else%
    \hss\egroup%
  \fi}%
\def\stablestart{%
  \ifstablesin%
    \errhelp=\stablesmultiplehelp%
    \errmessage{An S-Table cannot be placed within an S-Table!}%
  \fi
  \global\stablesintrue%
  \global\advance\stablescount by 1%
  \message{<S-Tables Generating Table \number\stablescount}%
  \begingroup%
  \stablestrutsize=\ht\stablestrutbox%
  \advance\stablestrutsize by \dp\stablestrutbox%
  \ifstablesborderthin%
    \stablesborderwidth=\stablesthinline%
  \else%
    \stablesborderwidth=\stablesthickline%
  \fi%
  \ifstablesinternalthin%
    \stablesinternalwidth=\stablesthinline%
  \else%
    \stablesinternalwidth=\stablesthickline%
  \fi%
  \tabskip=0pt%
  \stablesbaselineskip=\baselineskip%
  \stableslineskip=\lineskip%
  \stableslineskiplimit=\lineskiplimit%
  \offinterlineskip%
  \def\borderrule{\vrule width \stablesborderwidth}%
  \def\internalrule{\vrule width \stablesinternalwidth}%
  \def\thinline{\noalign{\hrule height \stablesthinline}}%
  \def\thickline{\noalign{\hrule height \stablesthickline}}%
  \def\trule{\omit\leaders\hrule height \stablesthinline\hfill}%
  \def\ttrule{\omit\leaders\hrule height \stablesthickline\hfill}%
  \def\tttrule##1{\omit\leaders\hrule height ##1\hfill}%
  \def\stablesel{&\omit\global\stablesmode=0%
    \global\advance\stableslines by 1\borderrule\hfil\cr}%
  \def\el{\stablesel&}%
  \def\elt{\stablesel\thinline&}%
  \def\eltt{\stablesel\thickline&}%
  \def\elttt##1{\stablesel\noalign{\hrule height ##1}&}%
  \def\elspec{&\omit\hfil\borderrule\cr\omit\borderrule&%
              \ifstablemode%
              \else%
                \errhelp=\stablelinehelp%
                \errmessage{Special ruling will not display properly}%
              \fi}%
  \def\stmultispan##1{\mscount=##1 \loop\ifnum\mscount>3 \stspan\repeat}%
  \def\stspan{\span\omit \advance\mscount by -1}%
  \def\multicolumn##1{\omit\multiply\stablestemp by ##1%
     \stmultispan{\stablestemp}%
     \advance\stablesmode by ##1%
     \advance\stablesmode by -1%
     \stablestemp=3}%
  \def\multirow##1{\stablesdummyc=##1\parindent=0pt\setbox0\hbox\bgroup%
    \aftergroup\emultirow\let\temp=}
  \def\emultirow{\setbox1\vbox to\stablesdummyc\stablestrutsize%
    {\hsize\wd0\vfil\box0\vfil}%
    \ht1=\ht\stablestrutbox%
    \dp1=\dp\stablestrutbox%
    \box1}%
  \def\stpar##1{\vtop\bgroup\hsize ##1%
     \baselineskip=\stablesbaselineskip%
     \lineskip=\stableslineskip%
     \lineskiplimit=\stableslineskiplimit\bgroup\aftergroup\estpar\let\temp=}%
  \def\estpar{\vskip 6pt\egroup}%
  \def\stparrow##1##2{\stablesdummy=##2%
     \setbox0=\vtop to ##1\stablestrutsize\bgroup%
     \hsize\stablesdummy%
     \baselineskip=\stablesbaselineskip%
     \lineskip=\stableslineskip%
     \lineskiplimit=\stableslineskiplimit%
     \bgroup\vfil\aftergroup\estparrow%
     \let\temp=}%
  \def\estparrow{\vfil\egroup%
     \ht0=\ht\stablestrutbox%
     \dp0=\dp\stablestrutbox%
     \wd0=\stablesdummy%
     \box0}%
  \def|{\global\advance\stablesmode by 1&&&}%
  \def\|{\global\advance\stablesmode by 1&\omit\vrule width 0pt%
         \hfil&&}%
  \def\vt{\global\advance\stablesmode by 1&\omit\vrule width \stablesthinline%
          \hfil&&}%
  \def\vtt{\global\advance\stablesmode by 1&\omit\vrule width
\stablesthickline%
          \hfil&&}%
  \def\vttt##1{\global\advance\stablesmode by 1&\omit\vrule width ##1%
          \hfil&&}%
  \def\vtr{\global\advance\stablesmode by 1&\omit\hfil\vrule width%
           \stablesthinline&&}%
  \def\vttr{\global\advance\stablesmode by 1&\omit\hfil\vrule width%
            \stablesthickline&&}%
  \def\vtttr##1{\global\advance\stablesmode by 1&\omit\hfil\vrule width ##1&&}%
  \stableslines=0%
  \stablesomitfalse}
\def\stablesdef{\bgroup\stablestrut\borderrule##\tabskip=0pt plus 1fil%
  &\stablesleft##\stablesright%
  &##\ifstablesright\hfill\fi\internalrule\ifstablesright\else\hfill\fi%
  \tabskip 0pt&&##\hfil\tabskip=0pt plus 1fil%
  &\stablesleft##\stablesright%
  &##\ifstablesright\hfill\fi\internalrule\ifstablesright\else\hfill\fi%
  \tabskip=0pt\cr%
  \ifstablesborderthin%
    \thinline%
  \else%
    \thickline%
  \fi&%
}%
\def\endtable{\advance\stableslines by 1\advance\stablesmode by 1%
   \message{- Rows: \number\stableslines, Columns:  \number\stablesmode>}%
   \stablesel%
   \ifstablesborderthin%
     \thinline%
   \else%
     \thickline%
   \fi%
   \egroup\stablesend%
\endgroup%
\global\stablesinfalse}
%

\noblackbox
\input epsf
\def\inbar{\vrule height1.5ex width.4pt depth0pt}
\def\IC{\relax\hbox{\kern.25em$\inbar\kern-.3em{\rm C}$}}
\def\IP{\relax{\rm I\kern-.18em P}}
\def\IF{\relax{\rm I\kern-.18em F}}
\def\IZ{\relax\ifmmode\hbox{Z\kern-.4em Z}\else{Z\kern-.4em Z}\fi}
\Title{\vbox{\baselineskip14pt
\hbox{DUKE-TH-96-106}\hbox{HUTP-96/A007}\hbox{hep-th/9602114}}}
{Compactifications of F-Theory on Calabi--Yau Threefolds -- I}
\bigskip\vskip2ex
\centerline{David R. Morrison}
\vskip2ex
\centerline{\it Department of Mathematics, Duke University}
\centerline{\it Durham, NC 27708, USA}
\vskip.25in\centerline{and}\vskip.25in
\centerline{Cumrun Vafa}
\vskip2ex
\centerline{\it  Lyman Laboratory of Physics, Harvard
University}
\centerline{\it Cambridge, MA 02138, USA}
\vskip .3in
We study compactifications of F-theory on certain Calabi--Yau threefolds.
We find that $N=2$ dualities of type II/heterotic
strings in 4 dimensions get promoted to $N=1$ dualities between
heterotic string and F-theory in 6 dimensions.  The
six dimensional heterotic/heterotic duality becomes a
classical geometric symmetry of the Calabi--Yau in the F-theory setup.
Moreover the F-theory compactification
sheds light on the nature of the strong coupling transition and what lies
beyond the transition
at finite values of heterotic string coupling constant.
\Date{\it {Final version, May~1996.}}

\newsec{Introduction}

\lref\phases{E. Witten, Nucl. Phys. {\bf B403} (1993) 159.}
\lref\agm{P.S. Aspinwall, B.R. Greene and D.R. Morrison,
Nucl. Phys. {\bf B416} (1994) 414; Nucl. Phys. {\bf B420} (1994) 184.}
\lref\deligne{P. Deligne, in: Modular Functions of One Variable IV,
Lecture Notes in Math. vol. 476, Springer-Verlag, Berlin (1975) 53.}
\lref\kas{A. Kas, Trans. Amer. Math. Soc. {\bf 225} (1977) 259.}
\lref\nakayama{N. Nakayama, in: Algebraic Geometry and Commutative
Algebra vol. II, Kinokuniya, Tokyo (1988) 405.}
\lref\kodaira{K. Kodaira, Annals of Math. {\bf 77} (1963) 563;
Annals of Math. {\bf 78} (1963) 1.}
\lref\kawamata{Y. Kawamata, J. Fac. Sci. Univ. Tokyo Sec. IA {\bf 30}
(1983) 1.}
\lref\fujita{T. Fujita, J. Math. Soc. Japan {\bf 38} (1986) 20.}
\lref\kmp{S. Katz, D.R. Morrison and M.R. Plesser, hep-th/9601108.}
\lref\asp{P.S. Aspinwall, hep-th/9510142; hep-th/9511171.}
\lref\km{A. Klemm and P. Mayr, hep-th/9601014.}
\lref\bsv{M. Bershadsky, V. Sadov and C. Vafa, hep-th/9510225.}

\nref\fvaf{C. Vafa, hep-th/9602022.}
\nref\ov{H. Ooguri and C. Vafa, Mod. Phys. Lett. {\bf A5} (1990) 1389;
 Nucl. Phys. {\bf B361} (1991) 469;
Nucl. Phys. {\bf B367} (1991) 83.}
\nref\olre{
M. Blencowe and M. Duff, Nucl. Phys.
{\bf B310}(1988) 387.}
\nref\huln{C. Hull, hep-th/9512181.}
\nref\nstu{P. Townsend
et. al., unpublished.}
\nref\MK{D. Kutasov
and E. Martinec, hep-th/9602049.}
\nref\ats{A. Tseytlin,
hep-th/9602064.}
\nref\ontd{E. Bergshoeff,  B. Janssenhep and T. Ortin,
hep-th/9506156.}
\nref\yan{R. Khuri and T. Ortin, hep-th/9512178.}
\nref\bars{I.
Bars, ``U-duality and M-theory'', talk presented at ``Frontiers in
Quantum Field Theory'', Osaka, Japan, Dec. 1995, to be published.}

In constructing compact examples of D-manifolds for
type IIB strings some evidence has emerged for the existence
of a 12 dimensional formulation of type IIB strings, the `F-Theory'
\fvaf.  The existence of a 12 dimensional viewpoint has been suspected
from various different viewpoints over the years
\refs{\ov,\olre}
and also more recently
\refs{\huln\nstu\MK\ats\ontd\yan{--}\bars}.
Let us briefly recall the setup in \fvaf .  The proposal
there was that there is a (10,2) theory underlying
the type IIB theory.  Upon compactifying 1 space and 1 time
coordinate, the physical degree of freedom of this compactification
is characterized by a complex structure $\tau$ of a torus.
Thus compactifications of F-theory can also be viewed as compactification
on manifolds which admit an elliptic fibration---in this way the compactified
manifold behaves {\it as if}\/ it is Euclidean, but via the map
discussed in \fvaf\ one can translate the geometry to that
of a (10,2) manifold.  In discussing compactifications
it is most natural to take this into account and consider
the manifold as if it is a Euclidean manifold with elliptic
fibers.  To be more precise one is considering an elliptically
fibered manifold {\it together}\/ with the choice of an embedded base
manifold.  Mathematically this means that we have an elliptically fibered
manifold together with a choice of a section\foot{We will
assume here that the base intersects the fiber at one point.  It would
be interesting to see whether or not multiple intersections
make sense.  If they do, they would correspond
already in 10 dimensions to type IIB-like theories which
have a $U$-duality group corresponding to subgroups of
$SL(2,\IZ)$.}.

Upon compactification of F-theory to 10 dimensions on $T^2$
we get type IIB theory.  Upon compactification on elliptic
$K3$ we get, as proposed in \fvaf, a model dual
to heterotic strings on $T^2$.  It is natural
to consider F-theory compactifications
on other manifolds, and the simplest next case is on a
Calabi--Yau threefold leading to $N=1$ theory in 6 dimensions.
 The most general manifold in this class to consider
is a Calabi--Yau threefold which admits an elliptic fibration.
We will divide these into two natural classes:  In the first
class we study elliptic Calabi--Yau manifolds which in addition
admit a $K3$ fibration; in other words we consider the case
where the $K3$ fiber itself is elliptically fibered.  This would
be useful for dualities with heterotic strings.  The next class
would be elliptic fibrations which do not admit a compatible $K3$
fibration.  (An example of this was studied in \fvaf---the model
there cannot be
dual to a heterotic string compactification as it has more
than one tensor multiplet.)  We consider the first case
in this paper.  The second type as well as certain aspects
of the first type will be discussed in a forthcoming paper \ref\mvt{
D.R. Morrison and C. Vafa, to appear.}.

The outline of this paper is as follows.  In section 2 we consider
general aspects of heterotic compactifications on $K3$ and further
compactification down to 4 dimensions on $T^2$.  We note some
of the type II duals proposed for these models \ref\kv{S. Kachru and C. Vafa,
Nucl. Phys. {\bf B450} (1995) 69.}\
and show how they lead to natural F-theory duals
already in 6 dimensions.   We also note the strong coupling
problem pointed out in \ref\dmw{M.J.
Duff, R. Minasian and E. Witten, hep-th/9601036.}\ in this context.
In section 3 we discuss certain mathematical facts for the elliptic Calabi--Yau
threefolds.
 In particular, just assuming that there is a dual F-theory
compactifications we can {\it derive}\/ the Calabi--Yau manifold needed
to compactify the dual F-theory; the manifolds we find agree with some
of those proposed in \kv .
Our derivation uses results
from algebraic geometry for elliptic manifolds
going back to the work of Kodaira \kodaira\ and others.
The present method goes
far beyond checking the spectrum of massless particles, and in fact
can be used as a systematic method to {\it construct}\/ type II
duals for heterotic string compactification on $K3$ or $K3\times T^2$.
We will see examples of this in the present paper, postponing a more
complete analysis to \mvt .

In section 4 we discuss the physics that these dualities teach us.
This will include geometrizing the heterotic/heterotic duality
recently proposed in \dmw\ as well as shedding light on
what lies beyond the strong coupling transition noted there.
It turns out that this transition gets mapped
on the F-theory side  to crossing a wall of the K\"ahler cone.  Luckily these
kind
of situations have been extensively studied in the context
of 2d conformal field theory and
mirror symmetry in \refs{\phases,\agm}.

There is some overlap between our work and a concurrently released
paper of Aspinwall
and Gross \ref\agnew{P.S. Aspinwall and M. Gross,
hep-th/9602118.}.

\newsec{Heterotic Compactifications on $K3$}

\nref\gs{M.B. Green, J.H. Schwarz
and P.C. West, Nucl. Phys. {\bf B254} (1985) 327.}
\nref\oet{J. Erler, J. Math. Phys. {\bf 35} (1994) 1819.}

In considering compactifications of the heterotic string on a manifold $M$
we have to choose
which gauge group ($SO(32)$ or $E_8\times E_8$) and what type of bundle
to use.  One requirement is that
the first Pontryagin number of the bundle
should be the same as that of the manifold:
$${1\over 2}p_1(V)={1\over 2}p_1(M) .$$
For example if we consider compactification on $M=K3$ and take
into account the fact
that half the Pontryagin number of $K3$ is 24, we learn that
we have to choose instanton number 24 for the gauge bundle.
An instanton number 24 gauge bundle generically breaks $SO(32)$
to $SO(8)$, so in this case for generic choices of gauge bundle
we expect an $SO(8)$ gauge symmetry in 6 dimensions.  In the case
of $E_8\times E_8$ the generic gauge symmetry we get depends
on how we distribute the instanton numbers $(k_1,k_2)$
among the two $E_8$'s.
If they are given by $(k_1,k_2)= (12,12)$ one can show \kv\
that generically
one does not obtain a gauge symmetry and the instantons
break both the $E_8$'s completely.   There are some other simple
cases that we consider for the sake of examples in this paper
including:
$$(k_1,k_2)=(24,0) \rightarrow G=E_8$$
$$(k_1,k_2)=(20,4) \rightarrow G=E_7$$
$$(k_1,k_2)=(18,6) \rightarrow G=E_6$$
$$(k_1,k_2)=(12,12) \rightarrow G= \{ 0 \} . $$
In all these cases we can determine which group we end up with
by Higgsing as much as possible as was done in \kv .
Moreover in the above examples for the resulting gauge group
$G$ we are left with no charged matter, which is somewhat special.
In all of the above cases, except for the last one,
there was a puzzle raised in \dmw:
$N=1$ theory in 6d is chiral and has potential anomalies.
The anomaly 8-form factorizes as
\refs{\gs,\oet}
$$I_8={1\over 16(2\pi)^4}(tr R^2 -v \,tr F^2)(tr R^2-{\tilde v}\,tr F^2)$$
Moreover it has been shown in \ref\Sag{A. Sagnotti,
Phys. Lett. {\bf B294} (1992) 196.}\
that this implies that the gauge kinetic term will
contain a term of the form
$${\cal L}\propto (v e^{-\phi} +\tilde v e^{\phi}) tr F^2 , $$
where ${\rm exp}(2\phi)=\lambda^2$ is the heterotic
string coupling constant in 6 dimensions.
In all the above examples---except the completely
Higgsed case of $(12,12)$ where $\tilde v=0$---one finds that $v$ and
$\tilde  v$ are both non-zero and have opposite
signs.  This in particular implies that for finite
values of heterotic string
coupling constant the gauge kinetic term becomes zero
at
\eqn\tpt{{\rm exp}(-2\phi)= {-\tilde v\over  v}}
suggesting a phase transition.  We will be able to shed
light on this phase transition once we construct the F-theory
duals of these heterotic vacua.  For later use let us list
the values of $\tilde v/v$ that we find for the cases above
$$SO(32)\rightarrow {-\tilde v\over  v}= 2$$
$$(24,0)\rightarrow {-\tilde v\over  v}=6$$
$$(20,4)\rightarrow {-\tilde v\over  v}= 4$$
$$(18,6)\rightarrow {-\tilde v\over  v}=3 . $$

\nref\vw{ C. Vafa and E. Witten, hep-th/9507050.}
\nref\KLM{A. Klemm, W. Lerche and P. Mayr,
Phys. Lett. {\bf B357} (1995) 313.}
\nref\kklmv{S. Kachru, A. Klemm, W. Lerche,
P. Mayr and C. Vafa, hep-th/9508155.}
\nref\witt{E. Witten, hep-th/9503124.}
\nref\hut{
C.M. Hull and P.K. Townsend, Nucl.
Phys. {\bf B438} (1995) 109.}
\nref\vun{C. Vafa, unpublished.}

Upon further compactification on $T^2$ we get a theory
in $d=4$ with $N=2$ supersymmetry.  For this
case there are a number of examples where a
dual string theory has been proposed corresponding to
type II compactification on Calabi--Yau threefolds.
For example, for the $SO(32)$ case the dual proposed is
\refs{\kv,\vw}
the Calabi--Yau
manifold defined by
$$M_{SO(32)}=(WP^{4}_{1,1,4,12,18},36) . $$
Among the other $E_8\times E_8$ examples mentioned above
for the first and the last one there were duals proposed in \kv;
a counting\foot{This counting was done jointly with Shamit Kachru.}
also suggests duals in the other two cases in terms of
Calabi--Yau's
which are $K3$
fibrations
\KLM :
$$M(24,0)=(WP^{4}_{1,1,12,28,42},84)$$
$$M(20,4)=(WP^{4}_{1,1,8,20,30},60)$$
$$M(18,6)=(WP^{4}_{1,1,6,16,24},48)$$
$$M(12,12)=(WP^{4}_{1,1,2,8,12},24) . $$
All of these proposed duals are hypersurfaces in weighted projective
spaces with the weights (as subscripts) and degree as given above.
The last example has been studied extensively
in the Coulomb phase in
\kklmv .

It is natural to expect that a duality between heterotic strings
and type II strings in 4 dimensions should lead to some statement in the
decompactification limit of $T^2$ and thus to a statement
about a 6-dimensional duality of strings.  In fact, as suggested in
\fvaf, there is such a duality in terms of F-theory.  Consider
the case where the manifold $M$ admits an elliptic fibration.
 Then type IIA on $M$ is on the same moduli as M-theory on
$M\times S^1$ and this in turn is on the same moduli as
F-theory on $M\times S^1\times S^1$.  So one would expect
that the six-dimensional heterotic string on $K3$ in the limit in
which $T^2$ decompactifies corresponds to the F-theory compactified
on $M$.  This can actually also be argued using adiabatic
arguments \vw :  Start from the compactification of F-theory on
elliptic $K3$ which is dual to heterotic compactification on
$T^2$.   Upon further compactification on $S^1$ we would
get the duality between M-theory on $K3$ to heterotic strings on $T^3$
\witt .
Upon compactification on $T^2$ we would get the duality between
type IIA strings on $K3$ and heterotic strings on $T^4$
\refs{\hut,\vun}.
However, we can do another thing:
We can consider
a one parameter family of the eight dimensional
dual theories, parametrized
by $\IP^1$, in such a way that on the $F$-theory side we get
a Calabi--Yau with an elliptic $K3$.  This automatically
implies that on the heterotic
side we get an elliptic $K3$ compactification with an
appropriate gauge bundle. Note that on the F-theory side
we have an elliptic fibration over a $\IP^1$-bundle over yet another $\IP^1$.
In other words the  base $B$ of the Calabi--Yau threefold
 is given by a $\IP^1$-bundle over $\IP^1$ with the fiber
being $T^2$.
Compactifications of this
system down from 6 dimensions to 5 and 4 on $S^1$
and $T^2$ respectively will give the chain of dualities suggested above.

So given the dualities proposed in \kv\ which admit
elliptic fibrations
 (see also other
examples in \ref\iet{G. Aldazabal,
A. Font, L.E. Ibanez and F. Quevedo, hep-th/9510093.}) we are led
to a conjectured duality between F-theory on the same
Calabi--Yau threefold and heterotic string on $K3$.

Let the Hodge numbers of the $K3$-fibered Calabi--Yau manifold
which also admits an elliptic fibration be given by $(h_{11},h_{12})$.
Let us count the multiplets in the 6 dimensional sense.
Since a tensor multiplet and a vector multiplet in $N=1$, $d=6$
will both lead to vector multiplets in $N=2,d=4$
and the $T^2$ compactification gives
rise to 2 additional vector multiplets, we learn that $r(V)+T=h_{11}-2$,
where by
$r(V)$ here we mean the rank of the vector multiplets
and by $T$ the number of tensor multiplets.
Moreover since the number of hypermultiplets are the same
in 6 and in 4 dimensions we learn that we have $h_{21}+1$ hypermultiplets
in 6 dimensions.  Out of these, $h_{21}$ (together with certain other
modes) correspond to complex moduli of Calabi--Yau.
It is a fact that the number of complex deformation of an elliptic
Calabi--Yau is the same as that of general Calabi--Yau in accordance
with this count of hypermultiplets.  (This
will be explained in the next section.)
The geometric origin of the last hypermultiplet will be discussed
below.

In the context of dualities with the heterotic string we know
that $T=1$, that is, there is only one tensor multiplet upon
compactification of the heterotic string on $K3$.  In particular
the scalar component of this tensor multiplet is the heterotic string
coupling constant.  Note that there are two K\"ahler modes
of F-theory on a manifold whose base is a $\IP^1$-bundle over
$\IP^1$, corresponding to the two K\"ahler classes of the $\IP^1$'s.
Let $k_f,k_b$ correspond to the K\"ahler classes of the fiber and
base $\IP^1$'s respectively.  Then the six dimensional
heterotic string coupling
constant is identified with
\eqn\hec{{1\over  \lambda^2}={\rm exp}(-2\phi)={k_b\over k_f}}
The other combination $k_fk_b$ is part of the extra
hypermultiplet left out in the count above.
To see \hec\ note that upon toroidal compactification to  eight dimensions
the heterotic string coupling constant $\lambda^2$ is identified
with $k_f$ \fvaf .  Since we are now considering a $\IP^1$ family
of them, and the six dimensional coupling gets rescaled as usual
by the volume $\lambda^2 \rightarrow \lambda^2/k_b$,
we find the relation \hec .

\newsec{Mathematical Aspects of Elliptic Calabi--Yau Threefolds}
In this section we discuss some mathematical
aspects of elliptic Calabi--Yau threefolds.  We will
mainly concentrate on the case where the base manifold
$B$ is a $\IP^1$-bundle over $\IP^1$; as explained
in the previous section this would be the class of interest
in constructing heterotic duals.  Some of our remarks
are however valid for arbitrary base manifold $B$.

In constructing heterotic duals the following observations
prove to be very crucial:  Suppose we know a heterotic string
has a gauge symmetry $G$ with no matter (for simplicity).
Then the proposed Calabi--Yau dual must have a singularity of type
$G$ \refs{\asp\bsv\km{--}\kmp}.
What this means, recalling the relation between
the elliptic fiber and the 7-brane, is the following:  The
regions where the elliptic modulus $\tau \rightarrow \infty$
correspond to the worldvolume of the 7-branes, which consists
of a surface $\Sigma$ sitting in the base $B$ together with the 6-dimensional
spacetime.  Depending on how the torus degenerates as we approach
$\Sigma \subset B$, we get various type of gauge groups
(characterized in the simplest cases of degeneration by A-D-E).
The regions where the torus degenerates may have
several components $\Sigma_i$ which would be identified
as (part of) several 7-brane worldvolumes.
Since the $c_1$ of the Calabi--Yau is zero it relates
a particular linear combination of the classes $[\Sigma_i]$
with the canonical class of the base $B$.
This is the generalization to the threefold
of the similar statement for elliptic $K3$'s.  In that case
the regions where the torus degenerates correspond
to points on the base $\IP^1$ and the condition for vanishing
$c_1$ when the fibration is generic is simply that we
have 24 of them.  (The condition is more complicated for non-generic
fibrations and will be discussed below.)
Similarly in the case of elliptic Calabi--Yau threefolds,
 the
linear combinations of the classes $[\Sigma_i]$
will depend on what type of singularities
$\Sigma_i$ correspond to.
 This will prove
very powerful in constructing the Calabi--Yau
duals for the heterotic models studied in the previous section
and allows one to have a constructive method for finding
the dual Calabi--Yau.

Note that effectively what we have learned via F-theory,
is that we {\it can}\/ talk about compactifications
of type IIB strings on certain manifolds with $c_1>0$,
and construct a compact version of D-manifolds \bsv , with
appropriate 7-brane skeletons arranged to cancel
the $c_1$.  This comment applies to the case of compactification
of F-theory in all dimensions; note that the worldvolume of the
7-brane intersects the base in complex codimension 1 which
is the correct dimension to cancel the $c_1$ of the manifold.
For instance for compactifications of F-theory on Calabi--Yau
4-fold (leading to $N=1$ in $d=4$)
the 7-brane skeleton inside the Calabi--Yau lives
on a 4 dimensional space, i.e. in complex codimension 1 on the base.

In this section we first discuss the relation between the classes
$[\Sigma_i]$ and the canonical class of the base.  We then talk about aspects
of
$\IP^1$-bundles
over $\IP^1$.  We then apply
this technology to construct the duals for the heterotic
side and recover the predictions of the previous section.
In the next subsection we discuss the condition of having
elliptic fibration for Calabi--Yau's.  It turns out
that not all the $K3$-fibered Calabi--Yau's admit an elliptic fibration.
Finally we explain why in the elliptic fibrations of threefolds
the number of complex deformations is the same as that
with relaxing the condition on elliptic fibration.

\subsec{Relation between the Geometry of the Base
and the $7$-Brane Worldvolume}

The types of singular fibers which can occur on an
nonsingular elliptic surface with no ``exceptional curves of the
first kind''\foot{Surfaces with $c_1=0$---the case of primary interest
for us---have no such ``exceptional curves of the first kind.''}
were classified long ago by Kodaira \kodaira.  All of
the components of these fibers are $\IP^1$'s (possibly with singularities),
and the ways in which they can be joined together are quite constrained.
For many of the fibers, the $\IP^1$'s are all nonsingular, and their
points of intersection all take the form of two $\IP^1$'s meeting
transversally.  In this case, it is convenient to represent the
fiber by means of a so-called dual graph whose vertices correspond to the
components of the fiber and whose edges indicate which components meet.
The list of such dual graphs is precisely the list of Dynkin diagrams
for the simply-laced affine Lie algebras, and we use the standard notation
for such diagrams (i.e., $\widetilde A_n$, $\widetilde D_n$, $\widetilde E_n$)
to indicate the type of fiber.
We display Kodaira's results in the first two columns of Table
1.\foot{We have omitted the cases on Kodaira's list which correspond
to so-called multiple fibers, since these do not occur when there is
a section of the fibration (as we are assuming).}
The fibers which are not associated to simply-laced affine Lie algebras
are displayed in Figure 1.

\bigskip
\begintable
Kodaira notation | Singular fiber | Weierstrass singularity   | $a_i$ \elt
I$_1$ | Fig. 1 | none | ${1\over 12}$ \elt
I$_{b}, b\geq 2$ | ${\widetilde A}_b$ | $A_b$ | ${b\over 12}$ \elt
II | Fig. 1 | none | ${1\over 6}$ \elt
III | Fig. 1 | $A_1$| ${1\over 4}$ \elt
IV | Fig. 1 | $A_2$ | ${1\over 3}$ \elt
I$^*_{b} ,b\geq 0$| ${\widetilde D}_{b+4}$ | $D_{b+4}$ | ${1\over 2}+{b\over
12}$
\elt
II$^*$ | ${\widetilde E}_8$ | $E_8$ | ${5\over 6}$ \elt
III$^*$ | ${\widetilde E}_7$ | $E_7$ | ${3\over 4}$ \elt
IV$^*$ | ${\widetilde E}_6$ | $E_6$ | ${2\over 3}$
\endtable

\centerline{\bf Table 1}

\let\picnaturalsize=N
\def\picsize{4.0in}
\ifx\nopictures Y\else{\ifx\epsfloaded Y\else\fi
\global\let\epsfloaded=Y
\centerline{\ifx\picnaturalsize N\epsfxsize \picsize\fi \epsfbox{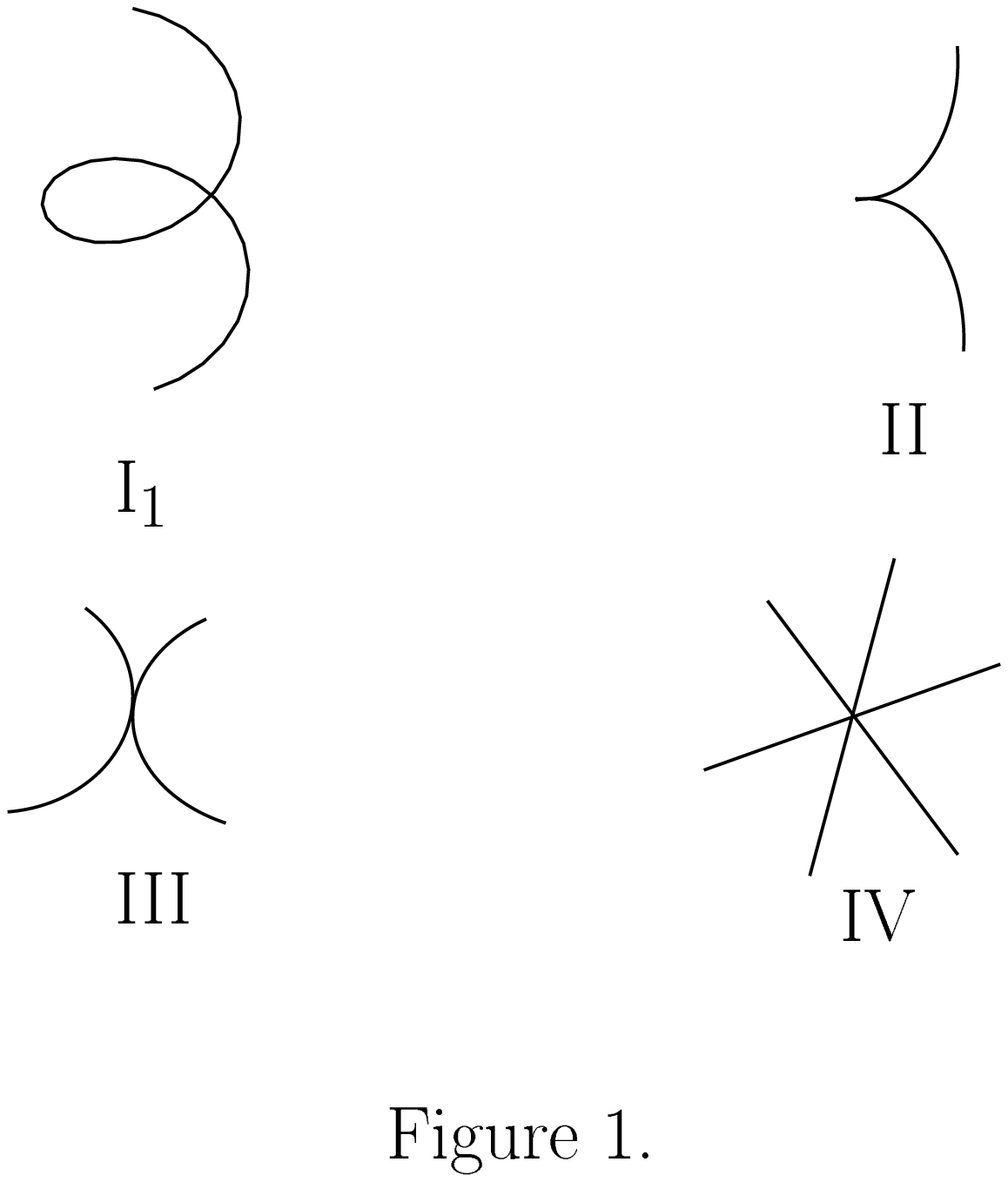}}}\fi

\lref\grassi{A. Grassi, Math. Ann. {\bf 290} (1991) 287.}

Kodaira found a formula for the canonical bundle of the total space $S$
which takes the form
\eqn\kodcan{
K_S=\pi^*(K_C+\sum a_iP_i),
}
where the sum is over points $P_i$ in the base $C$ over which the fibers
are singular, and the coefficients $a_i$ are determined by the type of
singular fiber as specified in the last column
of Table 1.  Thus, in order to obtain a $K3$ surface from this construction
we must have
\eqn\kthree{
K_C=-\sum a_iP_i.
}
This generalizes the ``generic'' case discussed above, since
if all of the singular fibers are of type I$_1$, then each coefficient
$a_i$ is equal to ${1\over12}$
and we find that there must be precisely 24 of them since $K_C$ has degree
$-2$.

\lref\aspkthree{P.S. Aspinwall, Phys. Lett. {\bf B357} (1995) 329.}

Since our fibration has a section, we can form a ``Weierstrass model''
by blowing down all components of each fiber which do not meet the
section \kas.  This is the geometric model we should expect to use in F-theory,
since there is only one ``size'' associated to the fiber.  And in fact,
the singularities which we introduce in this way are precisely of the
type (so-called A-D-E singularities) which should lead to gauge symmetry
enhancement in the type IIA theory compactified on
$K3$ \refs{\witt,\aspkthree}.  String/string duality
predicts that the gauge group should be the one whose Dynkin diagram
corresponds to the resolution graph of the singularity.  The
singularities we get in Weierstrass models are shown in the third
column of Table 1.

Kodaira's formula \kodcan\ was extended by Kawamata \kawamata, Fujita
\fujita, and Nakayama \nakayama\ to the
case of an elliptic fibration over a base $B$ of higher dimension,
where it takes the form
\eqn\kawcan{
K_M=\pi^*(K_B + \sum a_i[\Sigma_i]) + \hbox{\rm error term}.
}
The coefficients $a_i$ are again determined by the type of singular
fiber at the general point of each component $\Sigma_i$ of the locus
within $B$ on which the elliptic curve degenerates.  The error term
is present due to less accurate control over the birational geometry
of these spaces in higher dimension; however, thanks to work of Grassi
\grassi\ we need not concern ourselves with this error term in studying
elliptic Calabi--Yau threefolds if we choose our birational model
correctly.\foot{Grassi's results require us to
allow for the possibility of some mild singularities on the base $B$.
However,
these singularities will not be present in any of the examples discussed
in this paper.  (In particular, $B$ has a smooth model if the
threefold has a Weierstrass model.)}
In particular, in order to get $K_M$ to vanish we require:
\eqn\gracan{
K_B=-\sum a_i[\Sigma_i].
}

\nref\bkk{P. Berglund, S. Katz and A. Klemm, ``Extremal transitions
between dual $N=2$ string models'', to appear.}

As in the $K3$ case, there is a ``Weierstrass model'' obtained by blowing
down all components of fibers not meeting the section \nakayama\
(provided that
we start with a good birational model \grassi).  This Weierstrass model will
have curves of singularities, which should contribute an enhanced gauge
symmetry group to the type IIA theory.  In fact, the curves of
singularities
arising in Weierstrass models of elliptic fibrations
were precisely the geometric tool used in
\kmp\ to study enhanced gauge symmetry.\foot{The earlier method of
\asp\ based on $K3$ fibrations can also be used to study the models
of the present paper.}  As explained there,
in order to obtain models
without charged matter, the components of the locus $\Sigma$ must
not meet each other.  (Models with charged matter stemming from
intersection points of components of $\Sigma$ are discussed in
\refs{\bsv,\bkk}.)

\subsec{$\IP^1$-Bundles over $\IP^1$}

The set of possible $\IP^1$-bundles over $\IP^1$ can be described
in a very concrete way.  We begin with the base $\IP^1$ represented
in the form $\IC^2-\{(0,0)\}/\IC^*$.  That is, we let $\lambda\in\IC^*$
act on the homogeneous coordinates $(s,t)$ by $(s,t)\mapsto(\lambda s,
\lambda t)$; $\IP^1$ is the quotient space by that action.

In order to form a $\IP^1$-bundle, we represent the fiber in a similar
way, as the quotient by the action $(u,v)\mapsto(\mu u,\mu v)$
for $\mu\in\IC^*$ (restricting to $(u,v0\ne(0,0)$).
In order to get a nontrivial bundle structure,
we should take the homogeneous coordinates $u$ and $v$
of the fiber to transform as sections of a line bundle over the base
$\IP^1$.  In other words, under $\lambda$ these should transform
as $(u,v)\mapsto(\lambda^nu,\lambda^mv)$.  By changing the generators
of our $\IC^*\times\IC^*$ action and switching $u$ and $v$ if necessary,
we can assume that $m=0$ and $n\ge0$.  The resulting quotient space
is known as the minimal ruled surface $\IF_n$; these are known to be
all of the possible $\IP^1$-bundles over $\IP^1$.

When $n$ is even, this manifold is topologically a product of two
$S^2$'s.  The K\"ahler class on such a manifold can be specified
by the areas of the two $S^2$'s, leading to two K\"ahler parameters
$k_f$ and $k_b$.  Geometrically, $k_b$ is the coefficient of the divisor
$D_s=\{s=0\}$ which is the fiber of the $\IP^1$-bundle.  The other
generating class $k_f$ is trickier to compute, but turns out to
be the coefficient of $D_v+(n/2)D_s$, where $D_v=\{v=0\}$.
(The cohomology ring is
naturally generated by $D_s$, $D_v$, and $D_u=\{u=0\}$, with a
relation $D_u=D_v+nD_s$ and the intersection pairing determined by
$D_s\cdot D_s=0$, $D_s\cdot D_v=1$ and $D_u\cdot D_v=0$.
It follows that the self-intersection of $D_v+(n/2)D_s$ is $0$.)
We note for later use that the canonical bundle of $B=\IF_n$ is given by
$K_B=-2D_v-(n+2)D_s$.

A general K\"ahler class is now specified as $k_bD_s+k_f(D_v+(n/2)D_s)$.
The area of the divisor $D_v$ is then given by
\eqn\area{
\mathop{\rm area}(D_v)
=D_v\cdot\left(k_bD_s+k_f(D_v+({n\over2})D_s)\right)
=k_b-({n\over2})k_f .
}
The fact that this area must be positive leads to the inequality
\eqn\ineq{
{k_b\over k_f}\ge {n\over2} ,
}
which defines one of the boundaries of the K\"ahler cone of $\IF_n$.  The other
boundary is given by the condition $k_f\ge0$.

\subsec{Constructing the F-Theory Duals}

We now turn to the construction of elliptic Calabi--Yau threefolds
(with a section) over a base $B$ which is one of the surfaces $\IF_n$.
We wish to consider models with no charged matter, so we insist that
the components of the locus $\Sigma$ of degenerate elliptic curves
should not meet each other.

The divisor $D_v$ plays a special r\^ole on $\IF_n$ in that $D_v\cdot
D_v=-n$
which is  negative
 (when $n>0$); we also have $K_B\cdot D_v=n-2$.
If $D_v$ is not one of the components of $\Sigma$,
then by \gracan, $-K_B$ can be represented by a divisor with
positive coefficients whose components do not include $D_v$.
It follows that $-K_B\cdot D_v\ge0$, which implies that $n\le2$.

On the other hand, if $D_v$ {\it is}\/ one of the components of $\Sigma$
and has coefficient
$a_v>0$ in the canonical bundle formula \gracan,  then
$$-2=(K_B+D_v)\cdot D_v=((1-a_v)D_v+\hbox{other terms})\cdot
D_v=(1-a_v)D_v\cdot
D_v.$$
(The last equality follows from the fact that
 $D_v$ is disjoint from the other divisors appearing in the formula.)
This determines the value of $n$ as being $n=2/(1-a_v)$, if we know
the singularity type over $D_v$ (and hence the value of $a_v$).
Thus, to produce a theory with no gauge group we should work with $n\le2$,
while if we want $G=SO(8)$, $E_6$, $E_7$, or $E_8$ we find $a_v={1\over2}$,
$2\over3$, $3\over4$, or $5\over6$ so we should use
$n=4$, $6$, $8$, or $12$, respectively.

Each Weierstrass model we are seeking can be described as a hypersurface
within a bundle over $B$ \refs{\deligne,\kas,\nakayama}.
The fibers of this bundle will be weighted
projective
spaces\foot{It is more common in the mathematics literature to
represent the fibers as ordinary projective spaces, but the weighted
projective space representation is equivalent.}
 $WP^2_{1,2,3}$, which we represent in terms of homogeneous
coordinates $z$, $x$, $y$ (not all zero) with an action by $\nu\in\IC^*$ of
$(z,x,y)\mapsto(\nu z,\nu^2x,\nu^3y)$.
The Weierstrass equation will take the form
\eqn\CYeq{
y^2=x^3-f(s,t,u,v)\,xz^4-g(s,t,u,v)\,z^6,
}
where $f$ and $g$ depend on the coordinates $s$, $t$, $u$, $v$ of $B$.
As in our constructions of
bundles in the previous subsection, the coordinates $z$, $x$ and
$y$ should be taken as sections of some line bundles over the base,
and we can specify which bundles we are using by means of their
transformation properties under $(\lambda,\mu)\in\IC^*\times\IC^*$
(the group used to construct $\IF_n$).  By an appropriate change of basis
of $(\IC^*)^3$ we can assume that the action of $(\lambda,\mu)$ on $z$
is trivial.  Moreover, the actions on $x$ and on $y$ are constrained
by the fact that $y^2$ and $x^3$ both occur in the equation of the
hypersurface, so must transform the same way.  The upshot is that
the transformation by $(\lambda,\mu)$ must take the form
$$(z,x,y)\mapsto(z,\lambda^{2\alpha}\mu^{2\beta}x,
\lambda^{3\alpha}\mu^{3\beta}y).$$

It is convenient at this point to summarize our description in the
following way.  We have homogeneous coordinates $s,t,u,v,x,y,z$
on which $(\lambda,\mu,\nu)$ act with exponents as specified in the
following table:
\halign{\indent#\qquad\hfil&\hfil#\quad\hfil&\hfil#\quad\hfil&
\hfil#\quad\hfil&\hfil#\quad\hfil&\hfil#\quad\hfil&\hfil#\quad
\hfil&\hfil#\quad\hfil\cr
&$s$&$t$&$u$&$v$&$x$&$y$&$z$\cr
$\lambda$&$1$&$1$&$n$&$0$&$2\alpha$&$3\alpha$&$0$\cr
$\mu$&$0$&$0$&$1$&$1$&$2\beta$&$3\beta$&$0$\cr
$\nu$&$0$&$0$&$0$&$0$&$2$&$3$&$1$.\cr
}
\noindent
Our Calabi--Yau threefold is obtained by starting with these homogeneous
coordinates, removing the loci $\{s=t=0\}$, $\{u=v=0\}$, $\{x=y=z=0\}$,
taking the quotient by $(\IC^*)^3$, and restricting to the solution set
of \CYeq.

\nref\mp{D.R. Morrison and M.R. Plesser, hep-th/9508107.}

In fact, we have just given a description of this Calabi--Yau as a
hypersurface in a toric variety, and all of the powerful techniques of
toric geometry can be brought to bear on these examples.  (See \agm\
for a review in the physics literature.)  Alternatively, we can use this
data to give a linear sigma model description of the theory \phases.
{}From either point of view, there are conditions which must be satisfied
in order to get a Calabi--Yau manifold from this data (see \mp\ for
a summary in physical terms).  The first of these is that the monomial
$stuvxyz$ should have the same weight as terms appearing in the
equation \CYeq.  This condition immediately lets us solve for the
unknown exponents, and we find $\alpha=n+2$, $\beta=2$.  The remaining
conditions (which, in the language of toric geometry, state that a
certain polyhedron should be ``reflexive'') lead to the restriction
 $n\le12$; if we also demand that the components of $\Sigma$ be
disjoint from one another (so that there
is no charged matter) then we are limited to the cases
$n=0,1,2,3,4,6,8,12$.

Notice that we can generically use the second and third $\IC^*$'s to
set the values of $v$ and $z$ to be $1$.  This leaves us with 5 homogeneous
coordinates and a single $\IC^*$; in fact, we have mapped our Calabi--Yau
to a hypersurface of degree $6n+12$ in $WP^4_{1,1,n,2n+4,3n+6}$.
For the cases of $n=4,6,8,12$, this reproduces the conjectured heterotic
duals with gauge groups $SO(8)$, $E_6$, $E_7$ and $E_8$!\foot{
The case $n=3$ corresponds to an $SU(3)$ gauge symmetry with
no matter on the heterotic side.  It is very likely that
this is dual to the heterotic model with instanton numbers $(9,15)$,
as there is a branch of this model with generic unbroken $SU(3)$ \dmw .
The value of heterotic string coupling constant
where there is phase transition is also consistent with this
identification.}
Note that this also explains the coincidence observed
in \kv\ that there seems to be a chain of dualities obtained
by shifting the weights of the projective space
by multiples of $(0,0,2,4,6)$ (similar remarks appear
in \iet ).  Note that in the above we have also understood why
not all the multiples of $(0,0,2,4,6)$ appear (i.e. why there is a gap and a
bound for values of $n$).
The cases with $n=0,1,2$ involve some interesting features; we will discuss
them in the next section.

\subsec{The (3,243) Models}

The construction of Weierstrass models over $\IF_n$
which we have given above leads, for $n=0,1,2$, to
three families of Calabi--Yau threefolds with Hodge
numbers $(3,243)$, all equipped with $K3$ fibrations
as well as elliptic fibrations \nakayama.  On the other hand for
$E_8\times E_8$ heterotic strings
there are three classes of instanton numbers where complete
Higgsing is possible leading to the same Hodge numbers \refs{\kv,\iet,\dmw}:
$(12,12),(11,13),(10,14)$.  It is natural to try and match
the above choices of $n$ with these three cases.  It was conjectured
in \kv\ that the $(12,12)$ is dual to the Calabi--Yau given
by $n=2$ above.  Subsequently it was conjectured in \iet\
that $(10,14)$ also lies on the same moduli.  If these
conjectures are both true, one would thus expect that two
of the above choices of $n$ are in fact connected.  This turns
out to be the case.  As we will see below $n=0$ and $n=2$ are connected
and represent the same Calabi--Yau.

The case $n=2$ can be represented in terms of hypersurfaces of degree 24
in $WP^4_{1,1,2,8,12}$.
Of the $243$ complex structure moduli of this family,
only $242$ are realized as deformations of the equation of the
hypersurface.
In other words, there is one ``non-polynomial'' deformation,
reminiscent of the examples studied in \kmp.

In fact, as in \kmp, there is a natural locus of enhanced gauge symmetry
for these models: it corresponds to blowing down the curve $D_v$ in
$\IF_2$, and the corresponding surface lying above it in the Calabi--Yau.
Doing so produces a singular Calabi--Yau space with a genus 1 curve of
$A_1$ singularities.  The results of \refs{\bsv,\km,\kmp} then suggest that
we should see an enhanced $SU(2)$ gauge symmetry with 1 adjoint of matter
along this locus.  Indeed, to produce the singularities on the Calabi--Yau
space, we have had to restrict to codimension 1 in K\"ahler moduli,
but also to codimension 1 in complex structure moduli (the codimension
1 space of those complex structures which can be represented within
$WP^4_{1,1,2,8,12})$.

As in the case of the examples in \kmp, it is possible to see the full
complex structure moduli by using a complete intersection model rather
than a hypersurface model.  The space $\IF_2$ with the curve $D_v$
blown down to a point can be represented as a quadric hypersurface
in $\IP^3$ with
a singularity, given by, say
\eqn\quadsing{
\xi\eta+\zeta^2=0,
}
with the homogeneous coordinates being $(\xi,\eta,\zeta,\tau)$;
the polynomials $f$ and $g$ used to describe the Weierstrass model
can then be represented as polynomials in $\xi$, $\eta$, $\zeta$ and $\tau$
of degrees $8$ and $12$.  If we perturb \quadsing\ to
\eqn\quadsmooth{
\xi\eta+\zeta^2=c\,\tau^2,
}
then we get the ``non-polynomial'' deformations of the Weierstrass
model.

When $c\ne0$, \quadsmooth\ defines a surface isomorphic to
$\IF_0=\IP^1\times\IP^1$.  In fact, for generic moduli we have reproduced
a Weierstrass model over $\IF_0$, the $n=0$ case mentioned above.
Thus we see that the $n=0$ and $n=2$ cases of our construction actually
form part of the same family.    In the next section we connect
the cases with $n>0$ with strong coupling phase transition
in the heterotic string \dmw .
Note that this observation is consistent
with the recent observation in \ref\ibaan{G. Aldazabal, A. Font,
L.E. Ibanez and F. Quevedo, hep-th/9602097.}\ that
if we consider a special subspace of $ (10,14)$ heterotic vacuum,
the strong coupling puzzle in \dmw\ is avoided in
this case by Higgsing\foot{
One can also check, as in section 4 that the value of heterotic string coupling
constant where a problem is expected to develop is $1/\lambda^2 =n/2=1$
and is consistent with this picture.}.
In the above picture we have drawn, this is the same as making
a smooth deformation from the $n=2$ case to the more generic $n=0$ case by
deforming to the more generic moduli.

The $n=1$ case would appear to be different, however,
and it is natural to identify it with the $(11,13)$ $E_8\times E_8$
heterotic vacuum\foot{This is also supported
by the fact that, as in discussion
in section 4, one can check that the strong
coupling transition occurs at $1/\lambda^2=n/2=1/2$.}.  We can try to
use the methods of the previous section and map the Calabi--Yau to
a hypersurface in $WP^4_{1,1,1,6,9}$, but the Hodge numbers associated
to that space are (2,272).  In fact, our Calabi--Yau maps to a hypersurface
within $WP^4_{1,1,1,6,9}$ which is so singular that its Hodge numbers,
which are again $(3,243)$,
differ from that of the generic Calabi--Yau hypersurface in that space.

\subsec{Condition for Existence of Elliptic Fibration}

Before considering the implications of the above dualities
 let us note that we do not expect that all $N=2$ dualities
between type IIA and heterotic strings in 4 dimensions
 come from $N=1$ duality between F-theory and the heterotic string
in 6 dimensions. The reason for this is that some of the dualities
may come from first compactifying on $T^2$ and then using
special moduli of $T^2$ to get enhanced gauge symmetry
and then freezing these moduli by turning on such
gauge fields on the $K3$.  For example the first
main example in \kv\ was of this type. For such cases
we thus expect that there should be no elliptic fibration.
This is not in contradiction with the fact
that the corresponding Calabi--Yau manifold has a $K3$ fibration,
and that $K3$ admits elliptic fibrations.
  The reason for that
is that not all the $K3$'s admit elliptic fibrations and this
in particular prevents the first model considered in \kv\
from having an F-theory dual in 6 dimensions.

\nref\oguiso{K. Oguiso, Internat. J. Math. {\bf 4} (1993) 439.}

In fact, the conditions for a Calabi--Yau threefold to admit either
an elliptic fibration or a $K3$ fibration are quite easy to state \oguiso.
In order to have an elliptic fibration, there must be an effective
divisor $D$ such that (1) $D\cdot\Gamma\ge0$ for all curves $\Gamma$,
(2) $D^3=0$, and (3) $D^2\cdot F\ne0$ for some other divisor
$F$.\foot{We are omitting one condition here: in order to obtain an
elliptic fibration with a {\it section}\/ as in this paper, we need
to demand that $D$ and $F$ can be chosen so that $D^2\cdot F$
is a small number.}  In order to have a $K3$ fibration,
there must be an effective
divisor $\widehat D$ such that (1) $\widehat D\cdot\Gamma\ge0$ for all
curves $\Gamma$, and (2) $\widehat D^2\cdot F=0$ for all divisors
$F$ (which implies that $\widehat D^3=0$).
For a Calabi--Yau with both kinds of fibration, the two fibrations will
be compatible (i.e., the $K3$'s will themselves be elliptically fibered)
if $D^2\cdot\widehat D=0$.

\nref\twoparam{P. Candelas, X. de la Ossa, A. Font, S. Katz and
D.R. Morrison, Nucl. Phys. {\bf B416} (1994) 481.}

It is easy to see from the properties of
 the intersection ring of the first model of \kv\ that it
does not have divisors of the type required for an elliptic fibration.
That ring was calculated in \twoparam\ to be generated by classes $H$
and $L$ with intersection numbers $H^3=4$, $H^2L=2$, $HL^2=0$, $L^3=0$.
The only effective divisors whose triple self-intersection is $0$ are
multiples of $L$ (which
define the $K3$ fibration), and multiples of $3H-2L$.
But for the latter class,
$(3H-2L)\cdot\ell=-2$, where $\ell$ is the class of a particular $\IP^1$
on the Calabi--Yau; thus, this divisor does not meet part (1) of
the stated condition.

\subsec{Counting of Complex Structure of Elliptic Calabi--Yau's}

There is one major difference between the $K3$ and Calabi--Yau cases
of F-theory compactification which may appear rather surprising at
first sight.  In the $K3$ case, it is only at special values of the moduli
that an elliptic fibration structure can be found.  However, for
Calabi--Yau's, if one member of a family contains an elliptic fibration
then the general member of that family will also contain such a structure.
We want to explain how this comes about.

In both cases, we can characterize the existence of an elliptic fibration
by means of the existence of certain effective divisors in the space.
The cohomology class of this desired divisor would not move as we
vary the moduli, but whether that class can be represented by an
effective divisor {\it can}\/ change.  In the $K3$ case, this is
very likely to change because for generic moduli the class no longer
lives in $H^{1,1}$ but has acquired a component in the $H^{2,0}$ and
$H^{0,2}$ directions---it can no longer be represented by any divisor,
effective or not.  In the Calabi--Yau case, the change if it occurs
can not be for that reason.

\nref\wilson{P.M.H. Wilson, Invent. Math. {\bf 107} (1992) 561.}

The fact that the divisor we need persists at generic moduli in
the Calabi--Yau case is essentially a consequence of the fact
that not only do the divisor classes remain within $H^{1,1}$, but also
 \wilson\
the K\"ahler cone is constant for generic moduli. The K\"ahler cone can
shrink at special values of moduli, but when it shrinks,
it does so away from the locus $D^3=0$, and so no new elliptic fibration
is  introduced by the shrinking.

\newsec{Physical Implications of F-theory/Heterotic Duality}
In this section we discuss the implications of the F-theory/
heterotic dualities constructed above.

\subsec{The Strong Coupling Phase Transition}
It was pointed out in \dmw\ that for generic
compactifications of the heterotic string, at finite
values of the heterotic string coupling constant
there will be an infinitely strong coupling
for gauge fields.  Using the relation between the heterotic
string coupling constant and the ratio of $k_b/k_f$ \hec\
and using \tpt , we learn that the
expected singularities is when
\eqn\trp{{k_b\over k_f}={-\tilde v\over  v}}
However as shown in the previous section we know that
there is a bound for $k_b\over k_f$ for a rational ruled
surface:
\eqn\boun{{k_b\over k_f}\geq {n\over 2}}
where $n$ defines the rational ruled surface $\IF_n$ as described
in the previous section.  We are thus led to the identification
of
\eqn\iden{{-\tilde v\over  v}  ={n\over 2}}
It is easy to see that in all the examples
constructed above this is a correct identity.
In fact the relation between instanton numbers $(k_1,k_2)$ and ${\tilde v\over
v}$ seems to be
very simple.  For instanton number $(k_1,k_2)$ with
$k_1\geq 12$, we have
$${-2\tilde v\over  v}=k_1-12$$
This is true in all the above examples and we believe it to be
of general validity (this is also related to the observations
in \ref\witnew{E. Witten, hep-th/9602070.}).  We would
thus expect the simple identification
$$n=k_1-12$$
Using this, and the observations in
the previous section, we see that the manifold $M(k_1,k_2)$ dual
to $E_8\times E_8$
heterotic strings on $K3$
with instanton numbers ($k_1,k_2$) should be given as a hypersurface
$$M(k_1,k_2)\subset WP^4_{1,1,k_1-12,2k_1-20,3k_1-30},$$
of degree $6k_1-60$,
possibly with additional singularities.
(For the cases  $k_1=12,13,14$ see the discussion in the
previous section).  We conjecture that this identification is true
for all the allowed instanton numbers
(and not just the ones we have discussed explicitly).
In a sense we have {\it derived}
this using the duality of F-theory and heterotic strings in eight
dimensions and by using adiabatic arguments which suggests
that the Calabi-Yau manifold should be elliptically fibered
over $\IF_n$.  This is also
consistent with the well known-fact that there
is no elliptically fibered Calabi-Yau threefold over $\IF_n$
for $n>12$ \nakayama .
Note that in this list,
the heterotic string with instanton numbers $(16,8)$ also
appears and gives the same
manifold we have discussed for the $SO(32)$ heterotic string.  We thus
conjecture that $E_8\times E_8$ heterotic string on $K3$ with
instanton numbers $(16,8)$ is on the same moduli as heterotic
string  (or Type I string) with $SO(32)$ gauge group on $K3$.

The identification of the $n$ with $-2\tilde v/v$
implies that in all the above examples, except for the case of
$(12,12)$ compactification of heterotic string where $n=0$,
there is a bound for the heterotic string beyond which there
is a phase transition.  What can we say about this phase transition
using the duality we have found?

\nref\str{A. Strominger, Nucl. Phys. {\bf B451} (1995) 96.}
\nref\gms{B.R. Greene, D.R. Morrison and A. Strominger, Nucl. Phys. {\bf
B451} (1995) 109.}

Luckily this type of singularity in the K\"ahler moduli of Calabi--Yau
manifolds has been studied extensively before
\refs{\phases,\agm}.
However the singularities studied there are in the context of sigma
models.  But in 6 dimensions, F-theory compactifications strictly speaking
do not correspond to
perturbative  `string vacua' as different types of $(p,q)$
strings have been used in their construction.  Another
way of saying this is that the dilaton has been turned on and
there are points where it gives a large value of the coupling constant.
Moreover we have used non-perturbative U-dualities even to define
the F-theory vacuum.
This in particular
means that we are not necessarily
 justified in using sigma model techniques to study
F-theory compactifications.  However we will consider the following:
Compactify first on $T^2$, in which case we get an ordinary type IIA
compactification on the same manifold being dual to the heterotic
string on $K3\times T^2$.  Then we can use the sigma
model techniques to study these singularities.
What we learn from this investigation is that
the vacuum makes sense beyond this singularity
but it ceases to have an interpretation as a compactification
on a geometric manifold; in other words we lose
the manifold description, but we can effectively
talk about the vacuum beyond this point.  There are in general
a number of different phases of the theory which can be seen
from this sigma model analysis.  We expect that these different
phases correspond to different string compactifications---we
are studying this issue now in the examples discussed above.
 Moreover
we can also use mirror symmetry to construct
the type IIB duals which would be equivalent to these
compactifications where we now encounter complex degeneration
as opposed to K\"ahler degenerations.  This is nicer to study
because it implies that we can use the same mirror
manifold to study beyond the transition and we just have to change the complex
structure.  Thus we will find a geometric description for this
phase transition in this way.
At any rate from this description it is clear that
in the four dimensional version certain
new modes become massless as we are crossing the transition
point, just as was the case for the conifold point
\refs{\str,\gms}
and other types of singularities
\refs{\asp\bsv\km{--}\kmp}.
By considering the large volume limit of $T^2$ we can then get insight
directly into the six-dimensional transition.
We are currently studying this in detail in some of the models
above.  There is preliminary evidence of enhanced gauge
symmetries at the transition point;
the results will be reported in \mvt .

\subsec{Heterotic/Heterotic Duality}

\nref\dufet{M.J. Duff,
hep-th/9509106.}
\nref\ser{
G. Cardoso, G. Curio, D. Luest, T. Mohaupt and S.J. Rey, hep-th/9512129.}

Soon after the $N=2,d=4$ dualities were proposed in \kv\
it was pointed out by Klemm, Lerche and Mayr \KLM\ that there are other
additional symmetries on the heterotic side of the type of $S-T$
exchange symmetry, which is implied by these dualities
but remain unexplained. On the type II side, this
symmetry is a classical geometric symmetry of the manifold.
This is true for both of the
main examples considered in \kv .
These were further considered in
\refs{\dufet,\ser}.
One of these cases, the case of $(12,12)$ imbedding of instantons,
was recently studied in \dmw\ in which a strong/weak
self-duality in six dimensions was proposed.  Upon further compactification on
$T^2$
this in particular means, using
well-known facts \ref\duff{M. Duff, Nucl. Phys. {\bf B442}
(1995) 47.},
the $S-T$ exchange symmetry.  Given
that the four dimensional duality in \kv\ is already geometric
and given the correspondence we have found with the F-theory compactifications
on the same manifold, we see that already in 6 dimensions
we should be able to see the duality proposed in \dmw\ as
a geometric symmetry on the F-theory side.

This is essentially obvious given the construction we gave
in the previous section:   We found that the Calabi--Yau
threefold can be described as an elliptic fibration over $\IP^1\times \IP^1$
where the elliptic fibration is given (in affine coordinates) by
$$y^2=x^3-f(z_1,z_2)\, x-g(z_1,z_2)$$
where $(z_1,z_2)$ are the coordinates of the $\IP^1\times \IP^1$ and
$f$ is of degree 8 in each of the $z_i$ and $g$ is of degree
12 in each of them.  Note that we have 243 complex deformation parameters
given by $13 \times 13 +9\times 9-3-3-1=243$ defining
the coefficients of the polynomials up to $SL(2,\IC)\times
SL(2,\IC)$ action
and rescaling of $x,y$.
 Clearly there is an exchange symmetry
if we exchange the two $\IP^1$'s and at the same time
change the coefficients of the polynomials so that the
coefficient of $z_1^kz_2^l$ is exchanged with that of $z_1^lz_2^k$
in each of the terms.  Note that since the heterotic string coupling
constant is given by the ratio of the K\"ahler classes of these two
$\IP^1$'s \hec , exchanging them inverts the heterotic string coupling
constant.  We have thus geometrized the symmetry
observed in \dmw\ and at the same time understood
its action on the (complex part of) hypermultiplets.
It is amusing to see how some of the tests in \dmw\ will
come out here.  Let us do the simplest case and ask
on what subspace we will get an enhanced $SU(2)$ symmetry.
This could have two types of origins
on the heterotic side:  perturbative or non-perturbative
through small sized instantons \ref\wism{E. Witten, hep-th/9511030.}.  Given
the fact that
on the F-theory side the strong/weak duality is manifest, it suffices
to concentrate on one of them, say
the perturbative side.   For the heterotic side if we
are interested in getting an $SU(2)$ gauge symmetry perturbatively we
have to imbed the instanton number 12 bundle in an $E_7\subset E_8$,
and this gives us a space with 116 real parameters smaller.  Noting
that for us half of the hypermultiplet
space is geometrically realized as complex deformation parameters
(with the other half coming from modes such as turning on the
four form $A^+$) we should expect
(assuming the enhanced gauge symmetry occurs on the subspace
where these extra modes are zero) on a real codimension 58 subspace,
or complex codimension 29.

Note that $SU(2)$ gauge symmetry corresponding to the
perturbative symmetry of heterotic strings should come from
an unbroken part of the gauge symmetry in 10 dimensions.  This
in particular means that even in the 8 dimensional limit of $K3$ we
should see a gauge symmetry.  Let $(z_1,z_2)$ denote the coordinates
of the fiber and base respectively.  The above condition translates
to having two 7-branes characterized by positions in $z_1$
coming together.  In other words
the base $\IP^1$ would
correspond to part of a 7-brane worldvolume with an $A_1$ singularity
(the non-perturbative one will correspond to the fiber $\IP^1$
corresponding to a 7-brane with $A_1$ type singularity).
Let us count how big this space is.

The 7-brane worldvolume (where the elliptic curve
degenerates) is given by the vanishing locus of the
discriminant
$$\Delta = 4 f^3 - 27 g^2.$$
What we
want is for the discriminant to vanish to second order along some curve
described by $z_1 = \lambda$ for some constant $\lambda$.
$\Delta$ will vanish to second order along that curve if and only if both
$\Delta$ and $\partial \Delta/ \partial z_1$ vanish along $z_1 = \lambda$.
To get $\Delta$ to vanish we must have
$$  4 f(\lambda,z_2)^3 = 27 g(\lambda,z_2)^2 .  $$
To get that to happen for all $z_2$ we need $f(\lambda,z_2) = 3 h(z_2)^2$ and
$g(\lambda,z_2) = 2 h(z_2)^3$ for some polynomial $h(z_2)$ of degree 4.

To get $\partial \Delta/ \partial z_1$ to vanish along $z_1 = \lambda$ we need
$$  12 f(\lambda,z_2)^2 f'(\lambda,z_2) = 54 g(\lambda,z_2) g'(\lambda,z_2)  $$
where we are using $f'$ and $g'$ to denote derivative with respect to $z_1$.
Substituting the previous result we have
$$  108 h(z_2)^4 f'(\lambda,z_2) = 108 h(z_2)^3 g'(\lambda,z_2).  $$
The solution to this is $g'(\lambda,z_2) = h(z_2) f'(\lambda,z_2)$.

Now we count parameters, making Taylor expansions of $f$ and $g$ around
$z_1 = \lambda$.  The zeroth order terms are $f(\lambda,z_2)$ and
$g(\lambda,z_2)$,
polynomials of degree 8 and 12.  For a general Calabi--Yau in our family this
accounts for 9 + 13 = 22 of the complex parameters.  Similarly, the first
order terms are $f'(\lambda,z_2)$ and $g'(\lambda,z_2)$ and for the general
Calabi--Yau this accounts for 9 + 13 = 22 additional parameters.
However, along the locus of gauge symmetry enhancement, these two terms are
specified by $h(z_2)$ and $f'(\lambda,z_2)$, of degrees 4 and 8, accounting
for only 5 + 9 = 14 parameters.  There is one additional parameter given by
$\lambda$ (the location of the gauge symmetry enhancement).  So the
complex codimension in the full space is $44 - 15 = 29$, as expected.

Note that the strong/weak duality
of heterotic strings which is realized on the F-theory
side by the exchange of the base and fiber implies that the
singularity occurs at a particular point on the base $\IP^1$.
Since the base is `visible' to the heterotic side,
because of the 8 dimensional duality of F-theory with
heterotic strings on $T^2$ \fvaf\ we see that the dual
to perturbative gauge symmetries on the
heterotic side occurs at moduli  where there are singularities
of the bundle/$K3$ at particular points on the
base of the $K3$ on the heterotic side.  This is
in accord with the interpretation of them as
heterotic instantons of zero size \dmw .  Actually using
the above duality one can in principle
analyze a whole class of various singularities
corresponding to either $K3$ singularities and bundle singularities
and translate that into statements about geometric singularities
of the elliptic Calabi--Yau on the F-theory side.  This would be interesting
to develop further.

We would like to thank D. Anselmi, M. Bershadsky,
 R. Dijkgraaf, S. Kachru and R. Plesser for
valuable discussions.
The research of D.R.M. is supported in part by NSF grant DMS-9401447,
and that of
C.V. is supported in part by NSF grant PHY-92-18167.
\listrefs
\end